\documentclass[aps,prd,showkeys,preprint,12pt]{revtex4-1}
\usepackage[english]{babel}
\usepackage{amsmath,amsthm,amsfonts,amssymb}
\usepackage{dsfont}
\usepackage{graphicx}

\usepackage[T1]{fontenc}
\usepackage{amsmath,amsfonts,amssymb}
\usepackage[mathcal]{eucal}
\usepackage{mathrsfs}
\usepackage{latexsym}
\usepackage{bm}
\usepackage[all]{xy}
\usepackage{dsfont}
\usepackage{caption}
\usepackage{subcaption}
\usepackage{dsfont}
\usepackage{upgreek}
\newcommand{\sech}{\text{sech}}

\begin{document}

\title{Regular string-like braneworlds}

\author{J. E. G. Silva}
\email{euclides.silva@ufca.edu.br}
\affiliation{ Universidade Federal do Cariri(UFCA), Av. Tenente Raimundo Rocha, Cidade Universit\'{a}ria, Juazeiro do Norte, Cear\'{a}, CEP 63048-080 - Brazil
}
\author{W. H. P. Brand\~{a}o}
\email{waldohpb@gmail.com}
\affiliation{Universidade Federal do Cear\'a (UFC), Departamento de F\'isica, Campus do Pici, Fortaleza - CE, C.P. 6030, 60455-760 - Brazil}
\author{R. V. Maluf}
\email{r.v.maluf@fisica.ufc.br }
\affiliation{Universidade Federal do Cear\'a (UFC), Departamento de F\'isica, Campus do Pici, Fortaleza - CE, C.P. 6030, 60455-760 - Brazil}
\author{C. A. S. Almeida}
\email{carlos@fisica.ufc.br}
\affiliation{Universidade Federal do Cear\'a (UFC), Departamento de F\'isica, Campus do Pici, Fortaleza - CE, C.P. 6030, 60455-760 - Brazil}




\begin{abstract}
In this work, we propose a new class of smooth thick string-like braneworld in six dimensions. The brane exhibits a varying brane-tension and an $AdS$ asymptotic behavior. The brane-core geometry is parametrized by the Bulk cosmological constant, the brane width and by a geometrical deformation parameter. The source satisfies the dominant energy condition for the undeformed solution and has an exotic asymptotic regime for the deformed solution. This scenario provides a normalized massless Kaluza-Klein mode for the scalar, gravitational and gauge sectors. The near-brane geometry allows massive resonant modes at the brane for the $s$ state and nearby the brane for $l=1$.
\end{abstract}

\maketitle

\section{Introduction}

The braneworld paradigm brought new geometrical solutions for some of the most intriguing problems in physics, as the hierarchy problem \citep{rs1,rs2} and the origin of the dark energy \citep{cosmology} and the dark matter \citep{darkmatter}.
The warped geometry allows the bulk fields to propagate into an infinite extra dimension \cite{kehagias} and provides rich
internal structure for the brane \citep{Bazeia}. In five dimensions, the brane can be realized as a four dimensional domain wall, whose topological features guarantees the brane stability \cite{dewolfe,gremm,resonance}.

In the codimension-2, the vortex scenarios are a suitable source for an axisymmetric brane, known as a string-like braneworld \cite{Olasagasti,Gregory1}. By adding one more extra dimension to the Bulk spacetime, the correction to the Newtonian gravitational potential turns out to be smaller than in 5D \cite{Gherghetta}. Moreover, in the thin string-like brane limit, no fine-tunning between the Bulk cosmological constant and the brane tension is needed \cite{Gherghetta}. The conical behaviour nearby the brane also provides a mechanism for the brane cosmological problem \citep{Kehagias:2004fb}.

However, a global vortex geometry exhibits a mild singularity \cite{Cohen} and a regular local Abelian 3-brane which satisfies the dominant energy condition \cite{Giovannini} and a baby skyrmion brane \cite{Brihaye:2010nf} which was only accomplished numerically. In the thin string-like brane limit, a vacuum $AdS_6$ traps the massless modes of the bosonic and fermionic fields and provides a smaller correction to the gravitational potential \cite{Gherghetta,Tinyakov,Oda1,Liufermions}. By considering a vanishing Bulk cosmological constant, a braneworld with a supersymmetric scalar fields and a cigar shape was found \cite{cigaruniverse}. Some brane-core geometries were proposed considering more involved transverse spaces, such as the soliton cigar \cite{stringcigar}, the resolved conifold \cite{resolvedconifold}, the catenoid \citep{Torrealba}, an apple shapped manifold \cite{appleshapped}, the torus \cite{T2} and other.

In this article, we present a new class of smooth thick string-like model and explore some of its physical and geometrical features. The brane-tension varies inside the brane-core and attains the $AdS$ regime asymptotically. The inclusion of a deformation parameter enables a continuous flow from the thin string into a thick brane with a core structure. The deformation parameter modifies the core properties, such as the behaviour of the stress-energy components and the variation of the curvature inside the core. Among the thick string-like solutions, we analyzed the properties of two models: the first solution has a bell-shaped source satisfying the dominant energy and the second, whose source exhibits an exotic asymptotic behaviour. For the dynamics of bulk bosonic fields in these thick brane scenarios, the Kaluza-Klein modes reveal interesting near and far brane characteristics.

This work is organized as follows. In Sec. \ref{section2} we present the smooth and deformed solutions and study their source and geometry properties.  In Sec. \ref{section3}, the Kaluza-Klein modes for the scalar, vector gauge and gravitational fields are studied and their features discussed. In Sec. \ref{cp}, final remarks and perspectives are outlined.

\section{Smooth string-like geometries}
\label{section2}

\hspace{0.47cm}Consider a six-dimensional spacetime $M_6$ whose bulk dynamic is governed by the Einstein Hilbert action with bulk cosmological constant $\Lambda$, namely \cite{Olasagasti,Gregory1,Cohen,Oda1,Oda2,Liufermions,Gherghetta,Tinyakov,Giovannini}
\begin{equation}
S_g=\int_{M_6}\left(\frac{R}{2\kappa_6}+\Lambda+\mathcal{L}_{m}\right)\sqrt{-g}d^6x,
\end{equation}
where $\kappa_6 = 8\pi/M_6^4$, $M_6^4$ is the six-dimensional bulk Planck mass and $\mathcal{L}_m$ is the matter Lagrangian for the source of the geometry. A string-like braneworld is an axially symmetric bulk $M_6$ whose metric can be written as \cite{Olasagasti,Gregory1,Cohen,Oda1, Oda2,Liufermions,Gherghetta,Tinyakov,Giovannini}
\begin{equation}
ds_{6}^2=\sigma(r)\eta_{\mu\nu} dx^\mu dx^\nu+dr^2+ \gamma(r)d\theta^2,
\end{equation}
where $0 \leq  r \leq r_{max}$ and $\theta \in [0,2\pi$). In order to guarantee a smooth geometry at the origin, we have the regularity conditions \cite{Oda1,Oda2,Liufermions,Gherghetta,Tinyakov,Giovannini}
\begin{equation}
\sigma(0)=(\sqrt{\gamma(0)})'=1\text{ and }\sigma'(0)=0,
\end{equation}
where the primes denote derivatives with respect to $r$. For a 3-brane at the origin, we impose the condition \cite{Olasagasti,Gregory1,Cohen,Oda1,Oda2,Liufermions,Gherghetta,Tinyakov,Giovannini}
\begin{equation}
\gamma(0) = 0.\label{constraint}
\end{equation}

From the matter Lagrangian $\mathcal{L}_m$ we define the stress-energy tensor as an axisymmetric and static stress-energy tensor in the form \cite{Olasagasti,Gregory1,Cohen,Oda1,Oda2,Liufermions,Gherghetta,Tinyakov,Giovannini}
\begin{equation}
\textbf{T}=t_0(r)e_\mu \otimes e^\mu +t_r(r) e_r \otimes e^r + t_\theta(r) e_\theta \otimes e^\theta.
\end{equation}

Defining \textit{A}(\textit{r}) := $\sigma'/\sigma$, \textit{B}(\textit{r}) := $\gamma'/\gamma$, the bulk Einstein equation yields to the system of coupled equations
\begin{equation}
\label{braneeinsteinequation}
\frac{3}{2}A'+\frac{3}{2}A^{2}+\frac{3}{4}AB+\frac{1}{4}B^2+\frac{1}{2}B'=-\kappa_6(\Lambda + t_0(r)),
\end{equation}
\begin{equation}
\label{radialeinsteinequation}
\frac{3}{2}A^2+AB=-\kappa_6(\Lambda + t_r(r)),
\end{equation}
\begin{equation}
\label{angulareinsteinsequation}
2 A'+\frac{5}{2}A^2=-\kappa_6(\Lambda + t_\theta(r)).
\end{equation}
The equations \eqref{braneeinsteinequation}, \eqref{radialeinsteinequation} and \eqref{angulareinsteinsequation} form a complex system of coupled equations. For a global vortex source, the geometry exhibits a mild singularity \cite{Cohen}. For a local vortex, only numerical solutions are known \cite{Giovannini}.

The angular Einstein equation \eqref{angulareinsteinsequation} is a nonlinear non-homogeneous Riccati equation for $A$. For $A=B=-c=\sqrt{2\kappa_6(-\Lambda)/5}$, we obtain an $AdS_6$ vacuum solution describing a thin string-like braneworld \cite{Gherghetta}. For $\Lambda <0$, the most general vacuum solution is $A(r)=c\tanh (\lambda r +c_1)$, where $\lambda=\sqrt{-5\kappa_6 \Lambda/8}$ and $c_1$ is an integration constant. Nonetheless, this solution does not vanish asymptotically and then, it does not provide a localized gravitational massless mode.

\begin{figure}[htb] 
       \begin{minipage}[b]{0.48 \linewidth}
           \includegraphics[width=\linewidth]{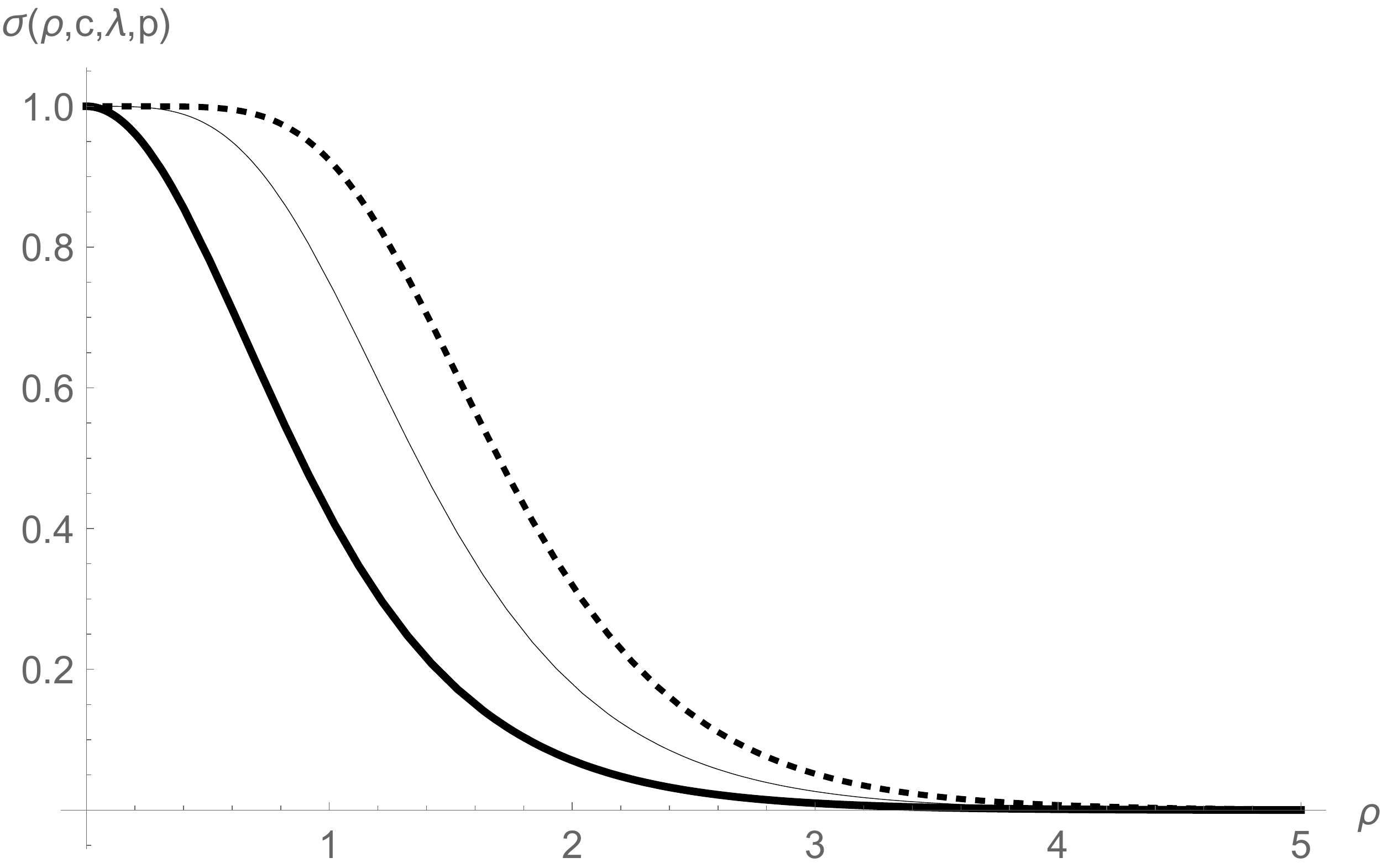}\\
           \caption{Warp factor for $p=1$ (thick line), $p=3$ (thin line) and $p=6$ (dotted line).}
          \label{deformed_warp_function}
       \end{minipage}\hfill
       \begin{minipage}[b]{0.48 \linewidth}
           \includegraphics[width=\linewidth]{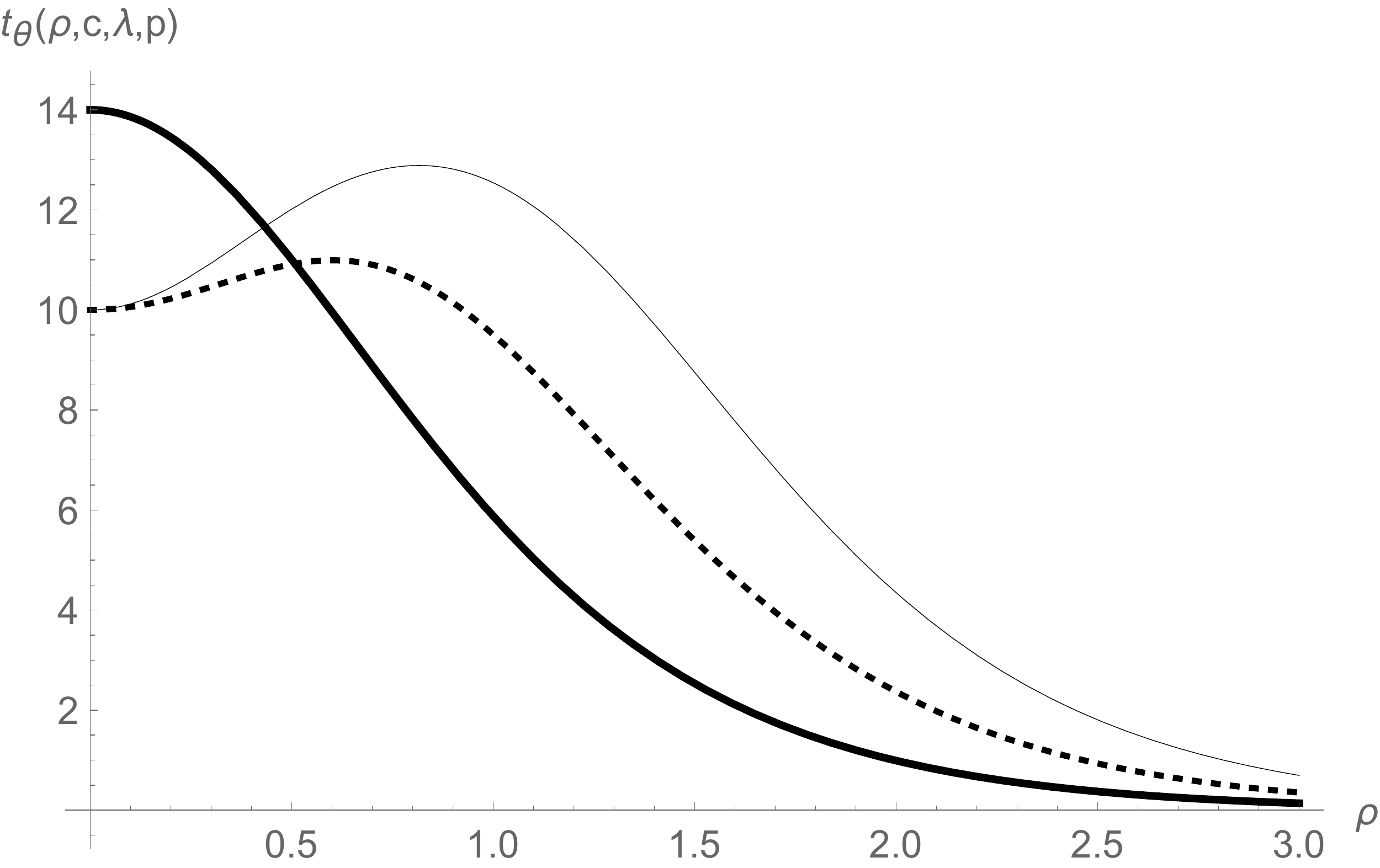}\\
           \caption{Angular pressure for $p=1$ (thick line), $p=3$ (thin line) and $p=6$ (dotted line). }
           \label{Fig-Espectro-Charuto}
       \end{minipage}
   \end{figure}

Since we are looking for thick smooth and localized geometries, let us assume an ansatz for the warp function $A(r)$ extending the thin string-like solution in the form
\begin{equation}
\label{warpfunctionansatz}
A_{c,p,\lambda}(r)=-c\tanh^{p}(\lambda r),
\end{equation}
where the parameter $c$ controls the asymptotic value of the warp function, $p$ modifies its variation inside the brane core and $\lambda$ determines the brane width. Note that for $p=0$ or for $p\neq 0$ and far from the origin  the warp function ansatz \eqref{warpfunctionansatz} reduces to the thin string-like solution. Therefore, the ansatz \eqref{warpfunctionansatz} represents a varying brane-tension solution.
Integrating Eq. \eqref{warpfunctionansatz} we obtain the warp factor $\sigma$ in terms of the hypergeometric function as
\begin{equation}
\label{deformedwarpfunction}
\sigma_{c,\lambda,p}(r)=\exp\left[-\frac{c \tanh^{p+1}(\lambda r)}{\lambda(p+1)} {}_2 F_{1}\left(1,\frac{p+1}{2};\frac{p+3}{2};\tanh^{2}(\lambda r)\right)\right].
\end{equation}
We plotted the warp factor \eqref{deformedwarpfunction} in Fig. \ref{deformed_warp_function}. For $p=1$ we obtain the bell-shaped $\sigma$
\begin{equation}
\sigma_{c,\lambda}(r)=\sech^{\frac{c}{\lambda}}(\lambda r),
\end{equation}
whereas for $p=3$ and $p=6$, the warp factor exhibits
a plateau around the origin. The deformed $p=2$ warp factor reduces to the string-cigar model \citep{stringcigar}.
For $p\in \mathbb{N}$, the source for the warp function $A$ has the angular pressure
\begin{equation}
t_\theta (r)=2cp\lambda \tanh^{p-1}(\lambda r)\sech^{2}(\lambda r) + \frac{5c^{2}}{2}\left(\sum_{i=1}^{p}(-1)^{i+1}\sech^{2i}(\lambda r) \right).
\end{equation}
The Fig. \ref{Fig-Espectro-Charuto} shows the behaviour of the angular pressure for $p=1$ (thick line), $p=3$ (thin line) and $p=6$ (dotted line). For $p=1$ the angular pressure has a bell-shape $t_\theta = (c(5c+4\lambda)/2)\sech^{2}(\lambda r)$ localized around the origin. For $p>1$,
the angular pressure exhibits two modified patterns inside the brane core. Therefore, we recognize the solutions for $p>1$ as deformed string-like branes.

The radial Einstein equation \eqref{radialeinsteinequation} provides a constraint between $A$ and $B$. Let us consider the ansatz for $B$ in the form
\begin{equation}
B(r)=A(r)+f(r).
\end{equation}
For $f=0$ and $c = 2\lambda$, the bulk metric has the form $ds^2=\sech^2(\lambda r)\eta_{\mu\nu} dx^\mu dx^\nu + dr^2 + R_{0}^2\sech^2(\lambda r)d\theta^2$ \cite{Torrealba}. Likewise the thin string-like model, this metric does not satisfy the condition expressed in Eq. \eqref{constraint}, and then, at $r = 0$, we have a 4-brane.

Since we seek for a regular geometry converging asymptotically to the $AdS_6$ spacetime and satisfying the regularity condition at the origin, let us adopt two models.

\subsection{Warped disk}
\begin{figure}[htb] 
       \begin{minipage}[b]{0.48 \linewidth}
           \includegraphics[width=\linewidth]{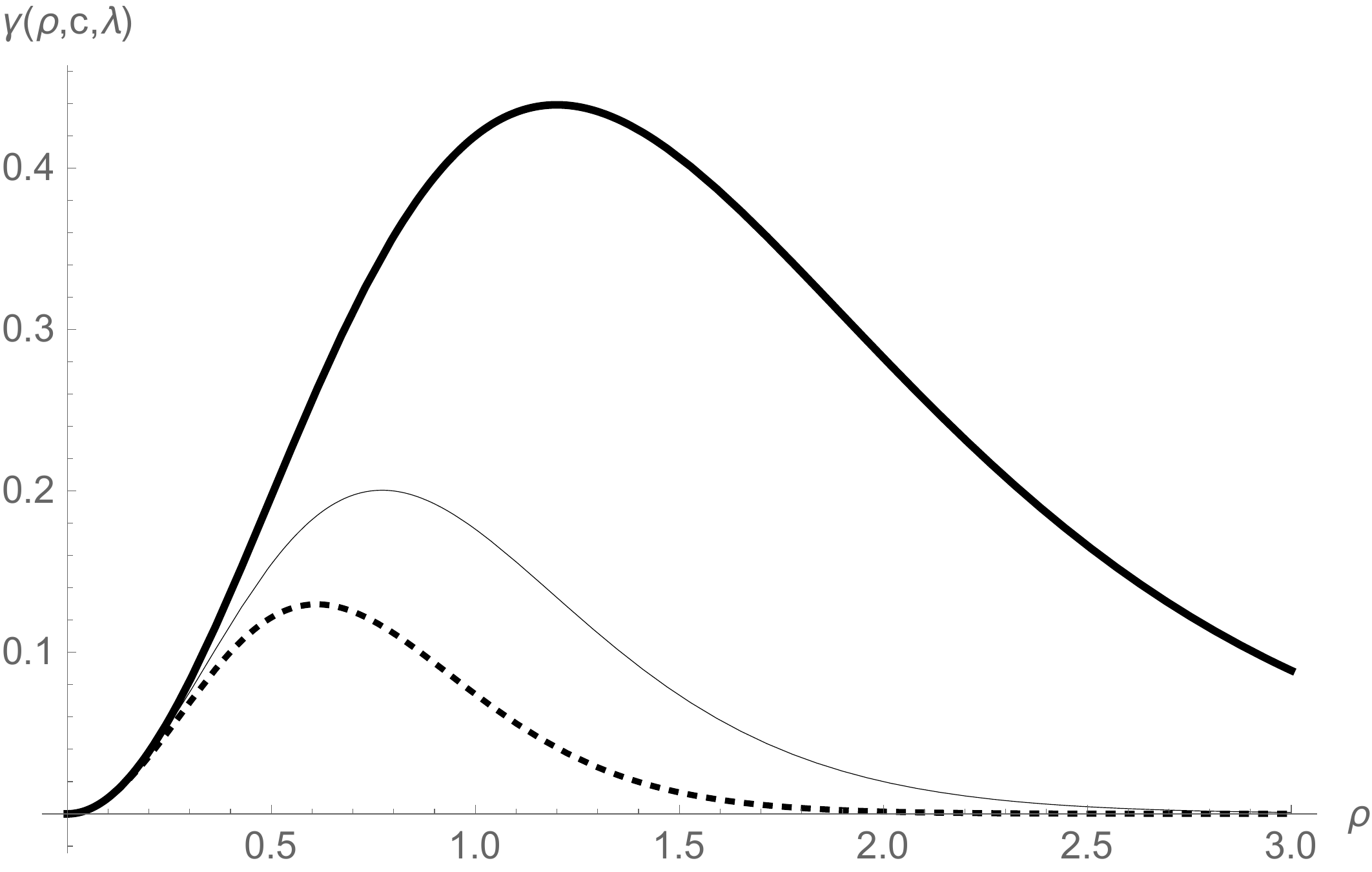}\\
           \caption{Angular metric component for $c/ \lambda=2$ (thick line), $c/ \lambda=4$ (thin line) and $c/ \lambda=6$ (dotted line).}
          \label{gamma_function_1}
       \end{minipage}\hfill
       \begin{minipage}[b]{0.48 \linewidth}
           \includegraphics[width=\linewidth]{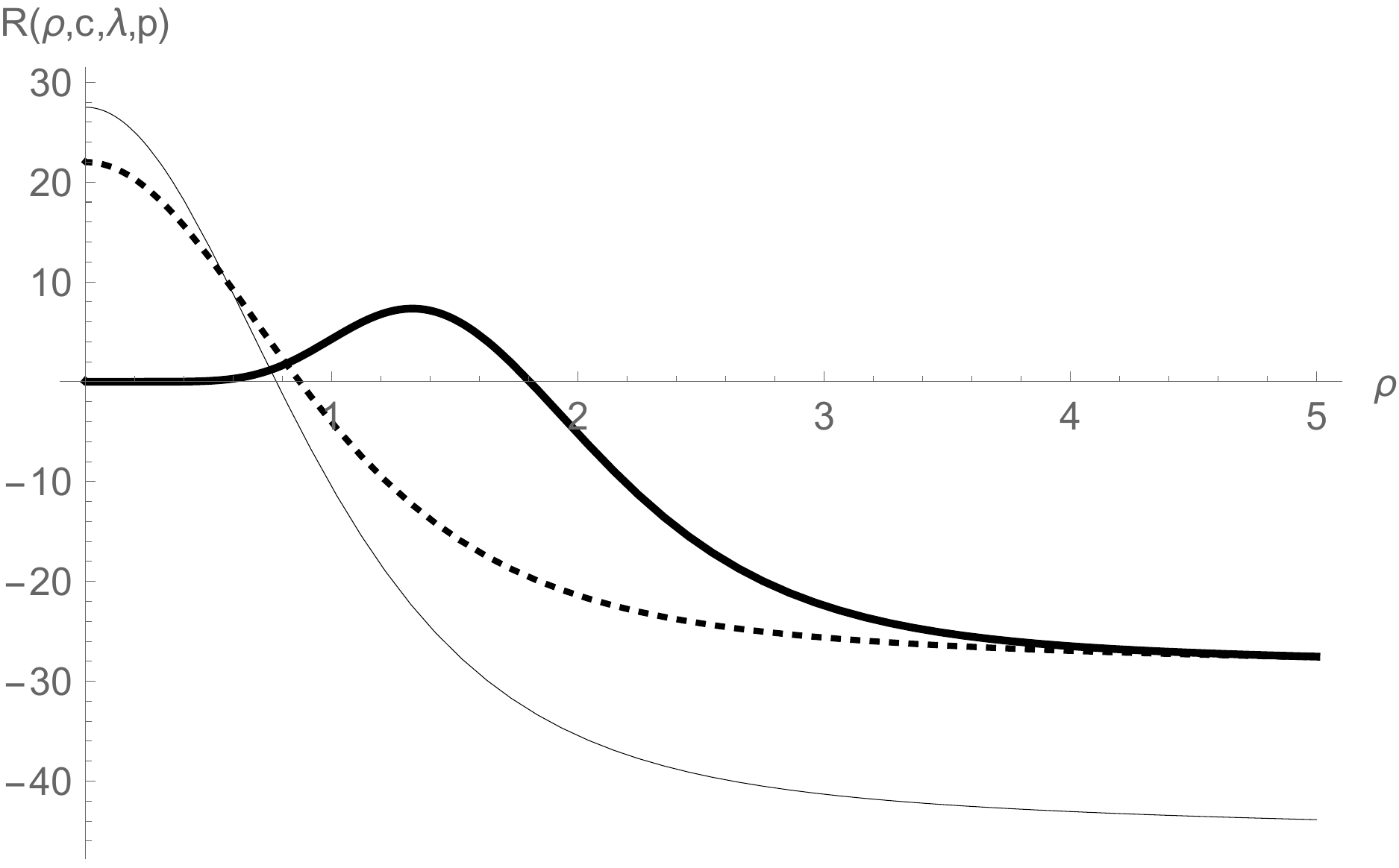}\\
           \caption{Ricci scalar for $p=1, c=2\lambda$ (dotted line), $p=1, c=2.5 \lambda$ (thin line) and  $c= 2\lambda$ and $p=10$ (thick line).}
           \label{scalar_curvature_disk}
       \end{minipage}
   \end{figure}

Consider $f_1(r):= \frac{m}{\lambda r}$. For this choice the metric has the form
\begin{equation}
\label{warpeddiskmetric}
ds^2=\sigma_{c,p,\lambda}(r)\eta_{\mu\nu} dx^\mu dx^\nu + dr^2 + r^{m/\lambda}\sigma_{c,p,\lambda}(r)d\theta^2.
\end{equation}
In order to satisfy the regularity conditions, we have to set $m=2\lambda$. Then, the metric \eqref{warpeddiskmetric} represents a warped product of the 3-brane and the two dimensional flat disk. The angular metric component is shown in Fig. \ref{gamma_function_1} for $p=1$ where the high of the function increases with the ratio $c/\lambda$. Fig. \ref{scalar_curvature_disk} shows the Ricci scalar which for $p=1$ has a bell-shape and for $p=10$ exhibits a plateau near the origin.  The components of the stress-energy tensor for $p=1$  are
\begin{eqnarray}
t_0 (r)&=&\kappa_6 \left(\frac{(5c+4\lambda)}{2}\sech^{2}(\lambda r)+\frac{5c}{2}\frac{\tanh(\lambda r)}{r}\right),\nonumber\\
t_r (r)&=&\kappa_6 \left(\frac{5c^{2}}{2}\sech^{2}(\lambda r)+2c\frac{\tanh(\lambda r)}{r} \right).
\end{eqnarray}

\begin{figure}[htb] 
       \begin{minipage}[b]{0.48 \linewidth}
           \includegraphics[width=\linewidth]{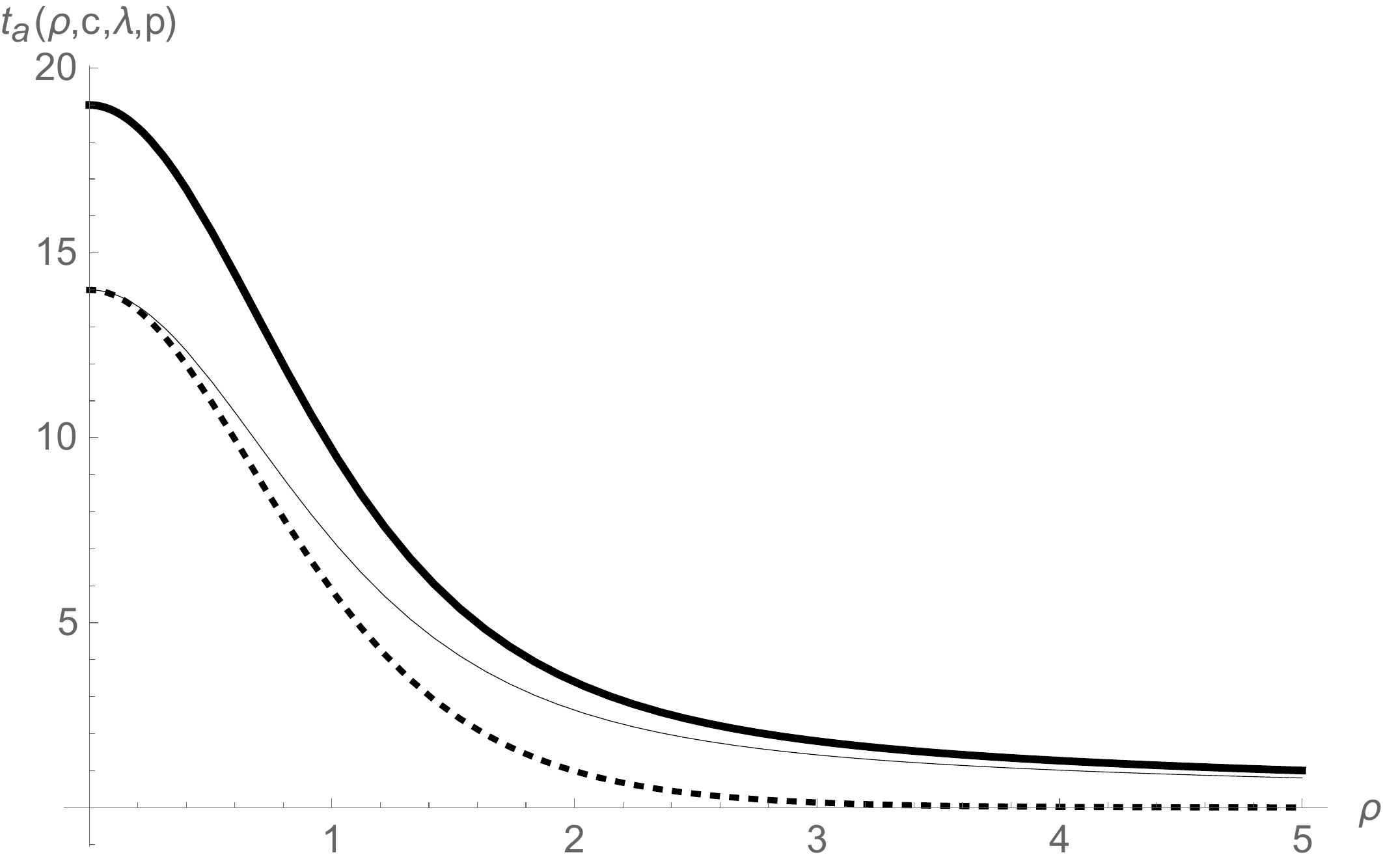}\\
           \caption{Energy density (thick line), radial pressure (thin line) and the angular pressure (dotted line) for $p=1$ and $c/ \lambda=2$.}
          \label{energy_components_disk}
       \end{minipage}\hfill
       \begin{minipage}[b]{0.48 \linewidth}
           \includegraphics[width=\linewidth]{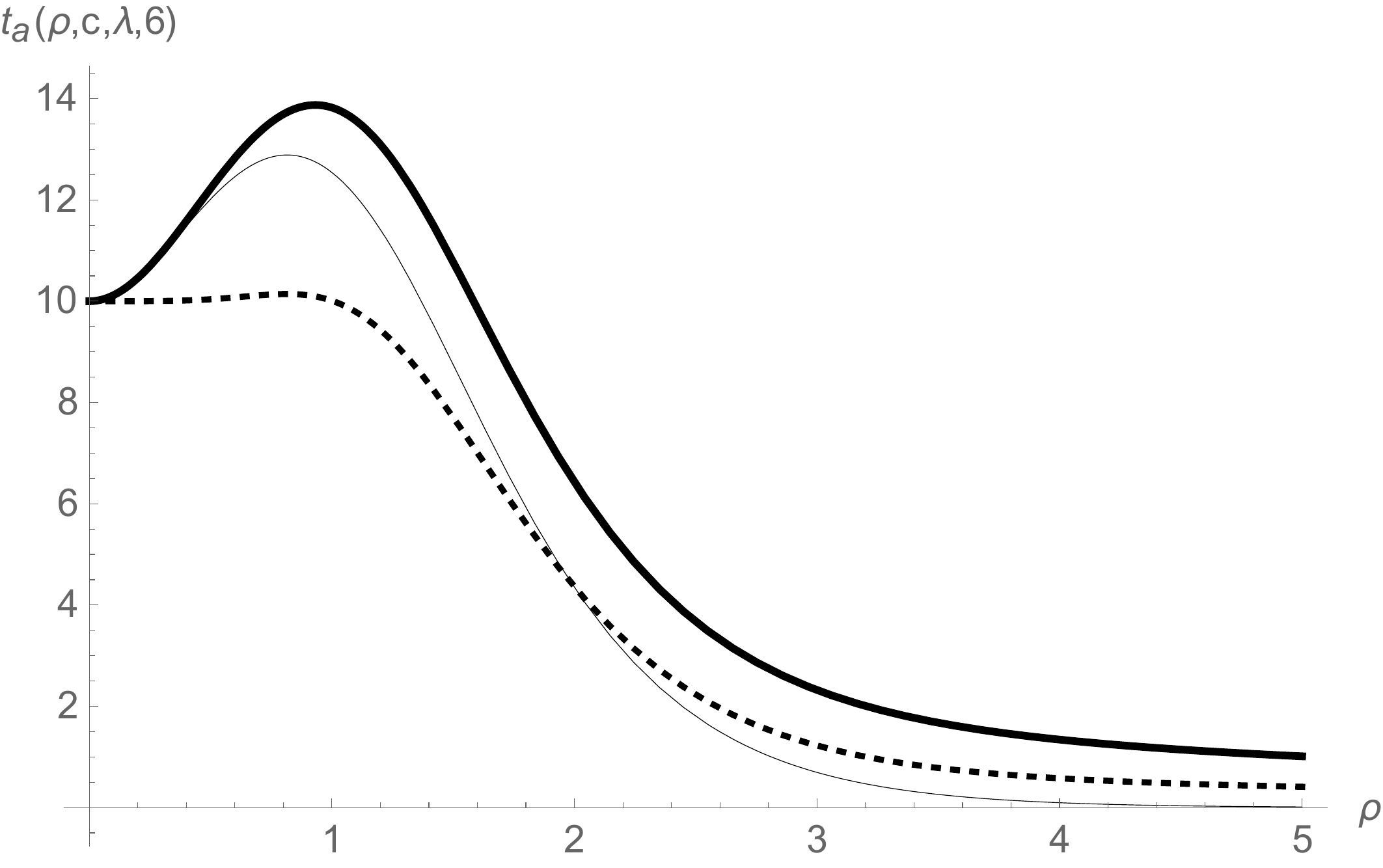}\\
           \caption{Energy density (thick line), radial pressure (dotted line) and the angular pressure (thin line) for $p=6$ and $c=2\lambda$.}
           \label{stress_energy_deformed}
       \end{minipage}
   \end{figure}

We plotted these components for $c=2\lambda$ in Fig. \ref{energy_components_disk}. Note that the source of this geometry is localized around the origin and satisfies the dominant energy condition. For $p=6$, the core shows an internal structure where the peak of the components are shifted from the origin and the radial pressure is almost constant.
The relationship between the bulk and brane Planck masses is given by
\begin{equation}
M_{4}^{2}=\frac{2\pi M_{6}^4}{\lambda} \int_{0}^{\infty}{r\sech^{\frac{3c}{2\lambda}}{\lambda r}dr}.
\end{equation}
For $c/\lambda=4/3$, we obtain $M_{4}^{2}=\frac{2\pi\ln{2}}{\lambda^{3}}M_{6}^4$. Then, the ratio $M_{4}/M_{6}$ increases as the brane width decreases.

\subsection{Exotic string-brane}

\begin{figure}[htb] 
       \begin{minipage}[b]{0.48 \linewidth}
           \includegraphics[width=\linewidth]{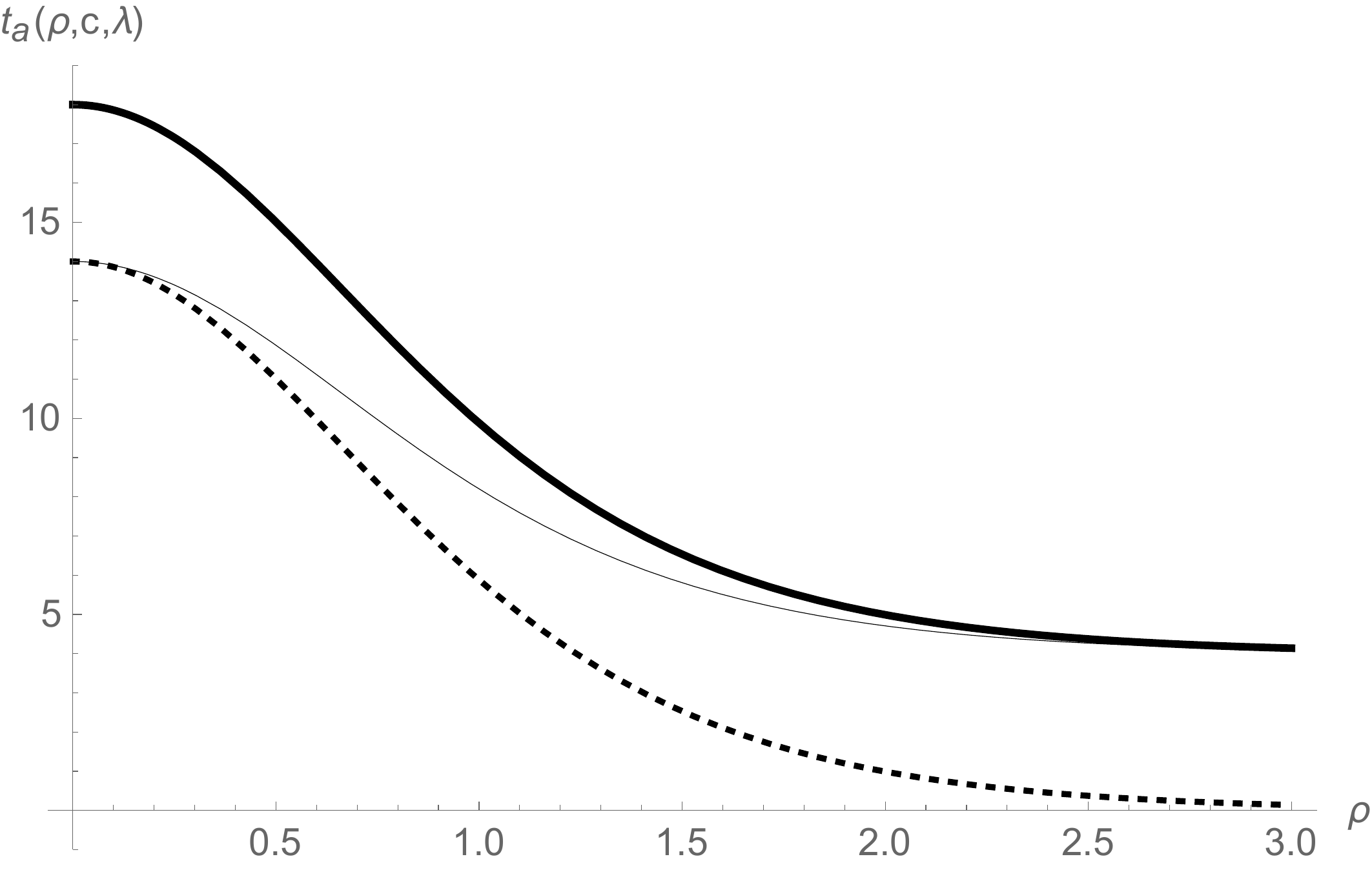}\\
           \caption{Energy density (thick line), radial pressure (thin line) and angular pressure (dotted line) for $p=\lambda=1, c=2$.}
          \label{energy_components_2}
       \end{minipage}\hfill
       \begin{minipage}[b]{0.48 \linewidth}
           \includegraphics[width=\linewidth]{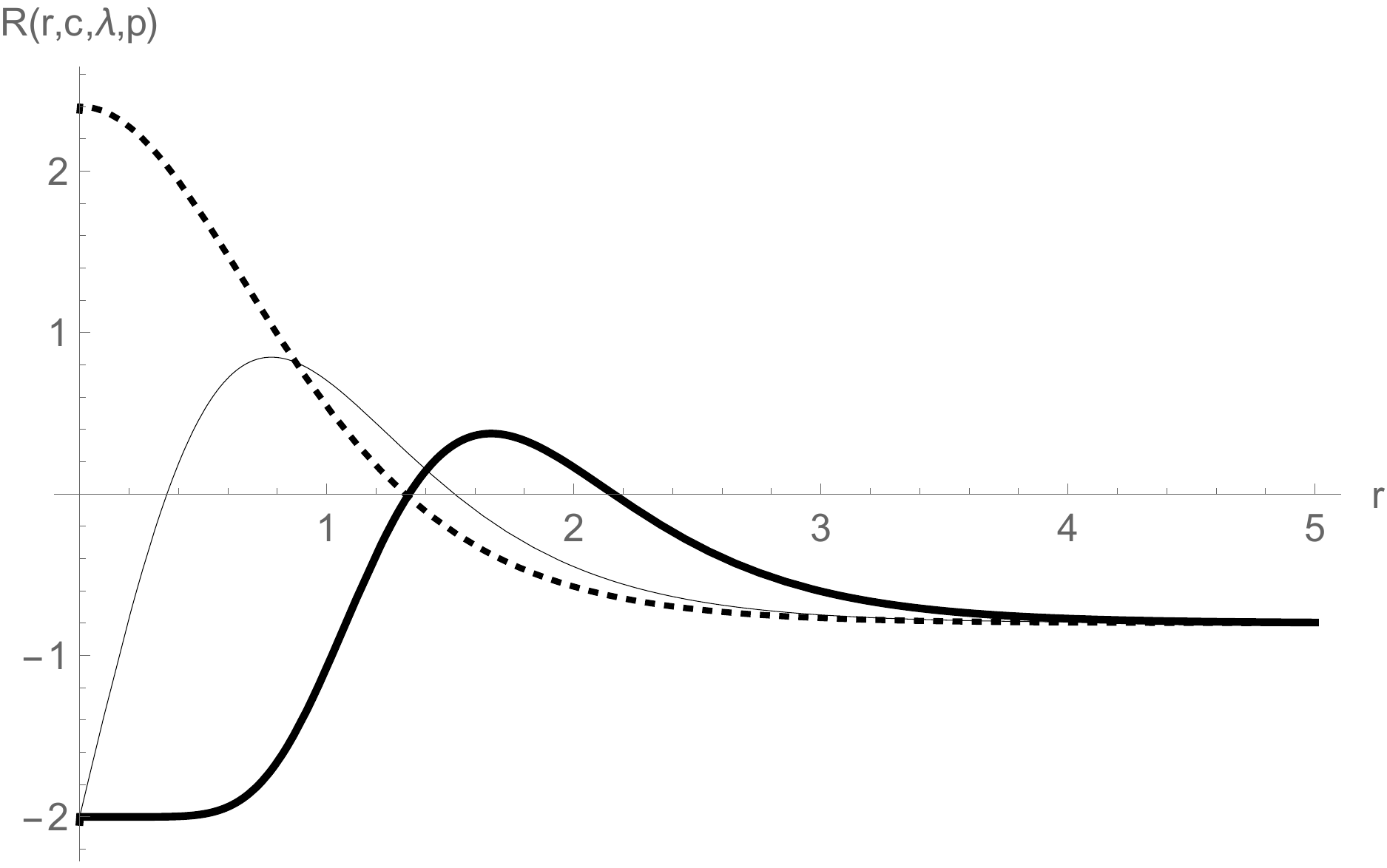}\\
           \caption{Ricci scalar for $c=\frac{2}{5},\lambda=1$ and $p=1$ (dotted line), $p=2$ (thin line), $p=10$ (thick line). }
           \label{scalar_curvature_exotic}
       \end{minipage}
   \end{figure}

For $f(r)$ = $\frac{m}{\tanh(\lambda r)}$,  the bulk metric has the form

\begin{equation}
ds^2=\sech^{c/\lambda}(\lambda r)\eta_{\mu\nu} dx^\mu dx^\nu + dr^2 + \frac{1}{\lambda^{\frac{m}{\lambda}}}\sinh^{\frac{m}{\lambda}}(\lambda r)\sech^{c/\lambda}(\lambda r)d\theta^2.
\end{equation}
The regularities conditions are satisfied for $m=2\lambda$. The components of the stress-energy tensor are
\begin{eqnarray}
t_0 (r)&=&\kappa_6 \left(\frac{c(5c+4\lambda)}{2}\sech^{2}(\lambda r)+\frac{\lambda(5c-2\lambda)}{2}\right),\nonumber\\
t_r (r)&=&\kappa_6 \left(\frac{5c^{2}}{2}\sech^{2}(\lambda r)+2c\lambda\right),
\end{eqnarray}
whose are sketched in Fig. \ref{energy_components_2}  for $c/\lambda=2$. The source satisfies the dominant energy condition inside the core and exhibits an exotic radial pressure asymptotically. For $c=\frac{2\lambda}{5}$, $t_0 = t_\theta$ and the source is dominated by the radial pressure.  Fig. \ref{scalar_curvature_exotic} shows the behaviour of the Ricci scalar as we change the deformation parameter $p$. For $p=1$, the curvature smoothly goes to an asymptotic $AdS_6$ spacetime whereas for $p=10$ there is a $AdS_6$ plateau near the origin.
In this scenario the bulk-brane relation mass is $M_{4}^2=\frac{4\pi}{(3c\lambda-2)\lambda^2}M_{6}^4,$
which increases as the source width $\lambda$ decreases.

\section{Bosonic fields\label{section3}}

\indent\indent In this section we study the effects of the warped disk and exotic string-like models have upon the gravitational, scalar and vector gauge fields.

\subsection{Gravity perturbations}
Assuming the metric perturbation $ds_6^2=\sigma(\eta_{\mu\nu} + h_{\mu\nu}(x^\zeta,r,\theta,k)dx^\mu dx^\nu +dr^2 + \gamma d\theta^2$, using the traceless transverse gauge $h_{\mu}^\mu=\nabla_{\mu} h^{\mu\nu}=0$ and performing the KK decomposition
$h_{\mu\nu}(x^\zeta,r,\theta)=\hat{h}_{\mu\nu}(x^\zeta)\sum_{l=0}^{\infty}\chi(r)e^{il\theta},$ the linearized Einstein equation yields to the radial graviton equation \cite{resolvedconifold}
\begin{equation}
\label{radialgravitonequation}
\chi''_{g}(r)+\left(\frac{5}{2}\frac{\sigma'}{\sigma}+\frac{1}{2}\frac{\beta'}{\beta}\right)\chi'_{g}(r)+\frac{1}{\sigma}\left(m^2-\frac{l^2}{\beta}\right)\chi_{g}(r)=0,
\end{equation}
where $m$ is the KK mass satisfying $\Box_4\hat{h}_{\mu\nu}=-m^2\hat{h}_{\mu\nu}$ and $\beta=\gamma/\sigma$. For a factorizable Bulk, i.e., $\sigma=1$ and $\gamma=r^2$, the gravitational KK solutions are $\chi_{g}=c_1 J_{l}(mr)+C_2 Y_{l}(mr)$. Thus, for a noncompact extra dimension the massive KK modes form a tower of non-normalized states. For a compact transverse space, the asymptotic behaviour of the
massive spectrum is $m_n \approx n\pi/R$. The warped scenarios bring a strong dependence of the cosmological constant and the brane source upon the KK modes, as we show in next sections.

\subsection{Scalar field}

Consider a minimally coupled massless scalar field $\Phi$ whose bulk action is given by
\begin{equation}
S_{\Phi}=-\frac{1}{2}\int{g^{MN}\nabla_{M}\Phi\nabla_{N}\Phi \sqrt{-G}d^{6}x}.
\end{equation}
The Bulk scalar field satisfies the equation of motion (EoM) 
\begin{equation}
\frac{1}{\sqrt{-G}}\partial_{M}(\sqrt{-G}g^{MN}\partial_{N}\Phi)=0,
\end{equation}
which after the Kaluza-Klein (KK) reduction $\Phi(x^\mu , r, \theta)=\hat{\Phi}(x^\mu)\sum_{l}{\chi_{s}(r)e^{il\theta}}$ yields to
\begin{equation}
\label{radialscalarfieldequation}
\chi_{s}''(r)+\left(\frac{5}{2}\frac{\sigma'}{\sigma}+\frac{1}{2}\frac{\beta'}{\beta}\right)\chi_{s}'(r)+\frac{1}{\sigma}\left(m^2-\frac{l^2}{\beta}\right)\chi_{s}(r)=0.
\end{equation}
It is worthwhile to stress that the graviton eq. \eqref{radialgravitonequation} has the same form of the radial equation for a massless scalar field \eqref{radialscalarfieldequation}. Accordingly, hereinafter we shall concern ourselves to the graviton radial function and the results extend to the scalar field straightforward.

\subsection{Vector field}
For the vector gauge field we assume a Maxwell action on the Bulk in the form
\begin{equation}
S_{A}=-\frac{1}{4}\int{g^{MR}g^{NS}F_{MN}F_{RS}\sqrt{-G}d^{6}x},
\end{equation}
where $F_{MN}:=\partial_{[M}A_{N]}$. Thus, the Bulk photon obeys the equation $\partial_{M}(\sqrt{-G}F^{MN})=0$ and assuming the gauge $\partial_{\mu}A^{\mu}=A_{\theta}=0$, the radial function $\chi_{A}(r)$ of the Kaluza-Klein decomposition $A_{\mu}(x^M)=\sum_{l=0}^{\infty}A_{\mu}^{(l)}(x^{\mu})\chi_{A}(r)Y_{l}(\theta)$ for $l=0$ satisfies \cite{gaugecigar}
\begin{equation}
\label{radialgaugeequation}
 \chi''_{A}(r)+\left(\frac{3}{2}\frac{\sigma'}{\sigma}+\frac{1}{2}\frac{{\beta}'}{\beta}\right)\chi'_{A}(r)+\frac{m^2}{\sigma}\chi_{A}(r)=0.
\end{equation}

It is worthwhile to stress the resemblance among the gravitational (scalar) and the gauge radial equations. In fact, for $l=0$, it is possible to unify these equation as
\begin{equation}
\label{radialunified}
 y''_{q}(r)+\left(\frac{q}{2}\frac{\sigma'}{\sigma}+\frac{1}{2}\frac{{\beta}'}{\beta}\right)y'_{q}(r)+\frac{m^2}{\sigma}y_{q}(r)=0,
\end{equation}
where $q=5$ for the gravitational and scalar fields and $q=3$ for the gauge vector field. In the following, we use the compact form in Eq.\eqref{radialunified} in order to compare the properties among the bosonic fields.

\subsection{Massless modes}
For $m=0$ (massless KK mode) and $l=0$ (s-state), the only normalizable solutions are the constants $\chi_{g0} =\chi_{s0}= N_g$ and $\chi_{A0} = N_A$. In the conformal variable $z=\int_{0}^{r}\sigma^{-\frac{1}{2}}dr'$, \textbf{these} gravitational \textbf{(scalar)} and gauge normalizable massless modes can be rewritten as 
\begin{eqnarray}
\Psi_{0g} & = &  N_g \sigma\beta^{\frac{1}{4}},\\
\Psi_{0A} & = &  N_A \sigma^{\frac{1}{2}}\beta^{\frac{1}{4}}.
\end{eqnarray}

\begin{figure}[htb] 
       \begin{minipage}[b]{0.48 \linewidth}
           \includegraphics[width=\linewidth]{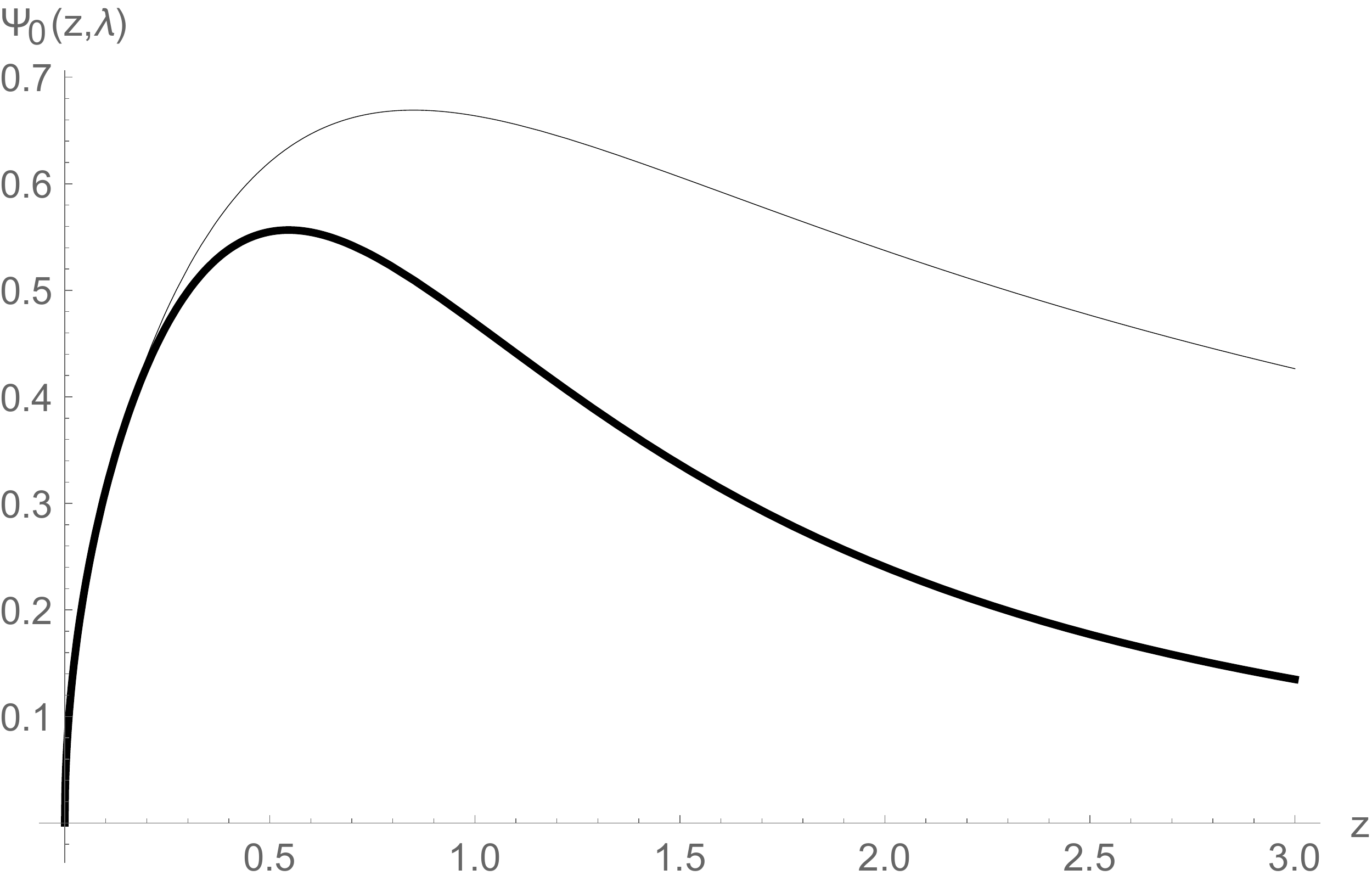}\\
           \caption{Gravitational (thick line) and vector gauge (thin line) massless modes for $l=0$ in the warped disk model $(c=2\lambda)$.}
          \label{zero_modes_1}
       \end{minipage}\hfill
       \begin{minipage}[b]{0.48 \linewidth}
           \includegraphics[width=\linewidth]{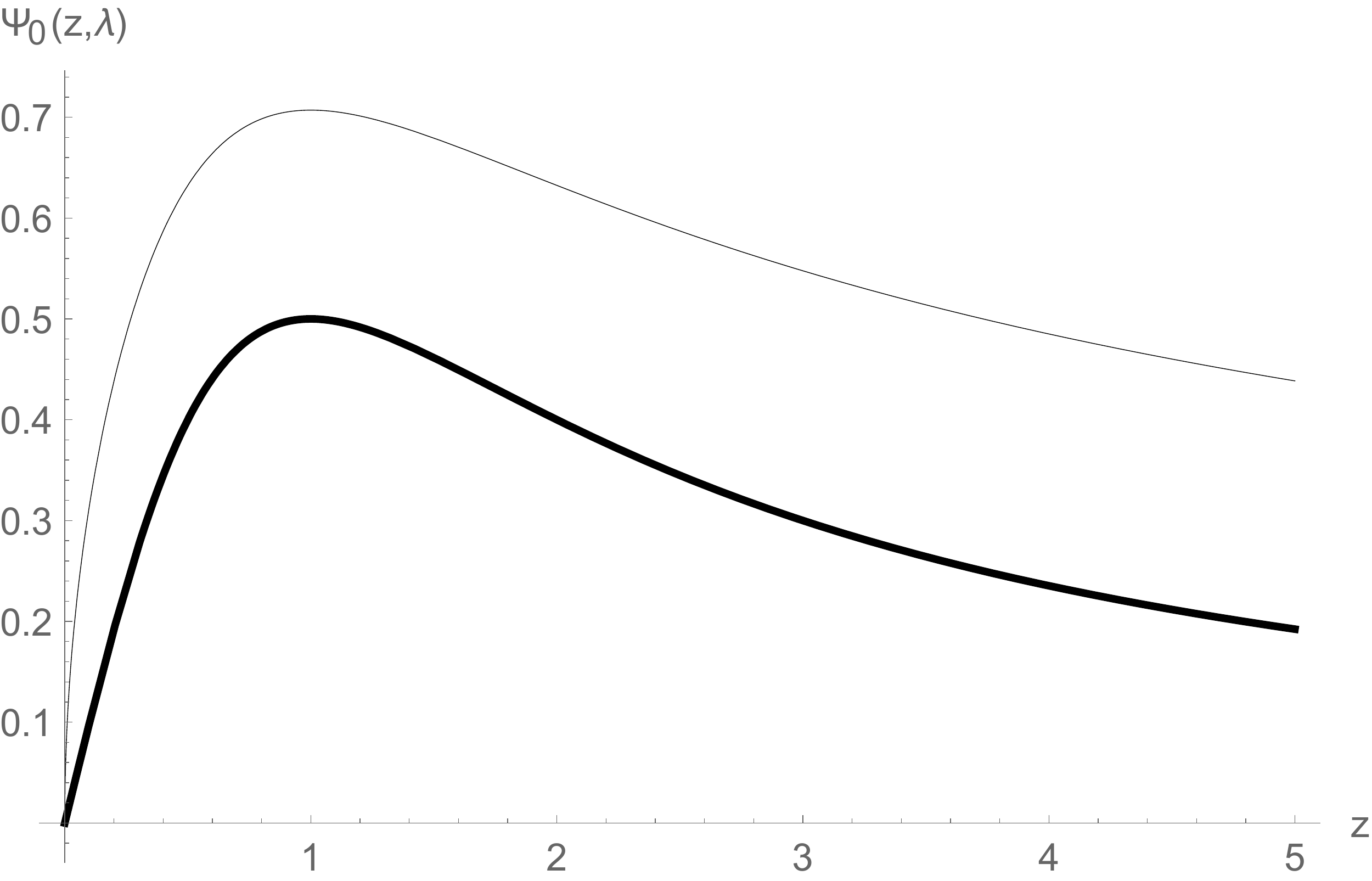}\\
           \caption{Gravitational (thick line) and vector gauge (thin line) massless modes for $l=0$ in the exotic string model $(c=2\lambda)$.}
           \label{zero_modes_2}
       \end{minipage}
   \end{figure}

For $c=2\lambda$, the conformal variable is given by $z=\sinh(\lambda r)/\lambda$ whereas for $c=6\lambda$, $z=(9\sinh(\lambda r)+\sinh(3\lambda r))/12$. For the warped disk model and $c=2\lambda$, the warp factor has the form $\sigma(z)=1/(1+\lambda^2 z^2)$ and $\beta(z)=(\sinh^{-1}{\lambda z})^2/\lambda^2$. We plotted the graviton \textbf{(scalar)} and photon massless modes for $\lambda=1$ in Fig. \eqref{zero_modes_1} and in Fig. \eqref{zero_modes_2}. Note that even though the massless modes are constant in the radial coordinate they exhibit a normalized profile in the conformal coordinate. The asymptotic exponential behaviour reflects the $AdS$ regime similar to the thin string-like scenarios \cite{Gherghetta}. The divergence near the origin stems from the change to the conformal coordinate.

\subsection{Massive modes}

For $m\neq 0$, the KK radial equation in the warped disk model becomes
\begin{equation}
\label{kksmoothdisk}
y_{m,l,q}^{''}+\left(\frac{1}{r}-\frac{qc}{2}\tanh(\lambda r)\right)y_{m,l,q}^{'}+\cosh(\lambda r)^{c/\lambda}\left(m^2 -\frac{l^2}{r^2}\right)y_{m,l,q}=0,
\end{equation}
where $q=5$ for the graviton (and the scalar field) and $q=3$ for the photon.
Near the brane, by expanding Eq.\eqref{kksmoothdisk} up to first order, we obtain
\begin{equation}
y_{m,l,q}^{''}+\left(\frac{1}{r}-\frac{qc\lambda r}{2}\right)y_{m,l,q}^{'}+\left(m^2 -\frac{l^2}{r^2}\right)y_{m,l,q}=0,
\end{equation}
whose solutions are
\begin{equation}
y_{m,l,q}(r)=A_{m,l,q}\sum_{k=0}^{\infty}{\Big[\frac{\left(qc\lambda l-2m^2\right)^k}{m^{2k}(2k)!}+(qc\lambda)^{k}\Big]\frac{(mr)^{2k+l}}{2^{2k}(l)_{k}}},
\end{equation}
where $A_{m,l,q}$ is an integration constant. The KK modes have a zero of order $l$ at the origin. Then, for $l=0$, $y_{m,0,q}(r)\rightarrow A_{m,l,q}$ as $r\rightarrow 0$, whereas for $l\neq 0$, $y_{m,l,q}(r)\rightarrow 0$. For $r<< \sqrt{2/qc\lambda}$ the KK modes are described by $y_{m,l,q}(r)=A_{m,l,q}J_{l}(mr)$. Hence, only the s-wave $l=0$ state is allowed on the brane. Asymptotically, Eq.\eqref{kksmoothdisk} becomes the thin string-like equation $y_{m,l,q}^{''}-\frac{qc}{2}y_{m,l,q}^{'}+m^2 \frac{e^{cr}}{2^{c/\lambda}}y_{m,l,q}=0,
$ whose solution is \cite{Gherghetta}:
\begin{equation}
y_{q,m,l}(r)=e^{\frac{qcr}{4}}\left(A_1 J_{q/2}\left(\frac{2^{\frac{2\lambda-c}{2\lambda}}me^{\frac{cr}{2}}}{c}\right)+A_2 Y_{q/2}\left(\frac{2^{\frac{2\lambda-c}{2\lambda}}me^{\frac{cr}{2}}}{c}\right) \right).
\end{equation}

For the exotic string model, the KK tower satisfies
\begin{equation}
\label{kkexotic}
y_{m,l,q}^{''}+\left(\lambda \coth (\lambda r) -\frac{qc}{2}\tanh(\lambda r)\right)y_{m,l,q}^{'}+\cosh(\lambda r)^{c/\lambda}\left(m^2 -\frac{(l\lambda)^2}{\sinh^2(\lambda r)}\right)y_{m,l,q}=0,
\end{equation}
which has the same near brane behavior of the warped disk model. However, far from the brane the KK eigenfunctions are given by
\begin{equation}
y_{q,m,l}(r)=e^{\frac{(qc-2\lambda)r}{4}}\left(A_1 J_{(qc-2\lambda)/2c}\left(\frac{2^{\frac{2\lambda-c}{2\lambda}}me^{\frac{cr}{2}}}{c}\right)+A_2 Y_{(qc-2\lambda)/2c}\left(\frac{2^{\frac{2\lambda-c}{2\lambda}}me^{\frac{cr}{2}}}{c}\right) \right).
\end{equation}

The presence of the width brane parameter $\lambda$ shows that the exotic source modifies the KK tower even outside the brane core.

Unlike the massless modes, the massive states forms a tower of non-normalizable states. Important qualitative features about the massive spectrum, as its stability and mass gap, can be obtained from the analogue Schr\"{o}dinger potential.
In the conformal coordinate $z=\int_{0}^{r}\sigma^{-\frac{1}{2}}dr'$ and defining
$\Psi_g = \sigma\beta^{\frac{1}{4}}\chi$ and $\Psi_A = \sigma^{\frac{1}{2}}\zeta^{\frac{1}{4}}$, we transform the Eq. \eqref{radialgravitonequation} and Eq. \eqref{radialgaugeequation} into a Schr\"{o}dinger-like equation \cite{Dantas}
\begin{equation}
-\ddot{\Psi}(z)+U(z)\Psi(z)=m^2\Psi(z),
\end{equation}
where $U_g (z)=\dot{W}_g (z)+W_{g}^2+\frac{l^2}{\beta}$, and $U_A (z)=\dot{W}_{A}(z)+W_{A}^2$ are the analogue Schr\"{o}dinger potential
and $W_{g}(z)=\frac{\dot{\sigma}}{\sigma}+\frac{1}{4}\frac{\dot{\beta}}{\beta}$, and $W_{A}(z)=\frac{1}{2}\frac{\dot{\sigma}}{\sigma}+\frac{1}{4}\frac{\dot{\beta}}{\beta}$ are the respective gravitational and gauge super potential \cite{Dantas}. The existence of such super potentials guarantees the absence of tachyonic (unstable) massive states \cite{Dantas}.

\begin{figure}[htb] 
       \begin{minipage}[b]{0.48 \linewidth}
           \includegraphics[width=\linewidth]{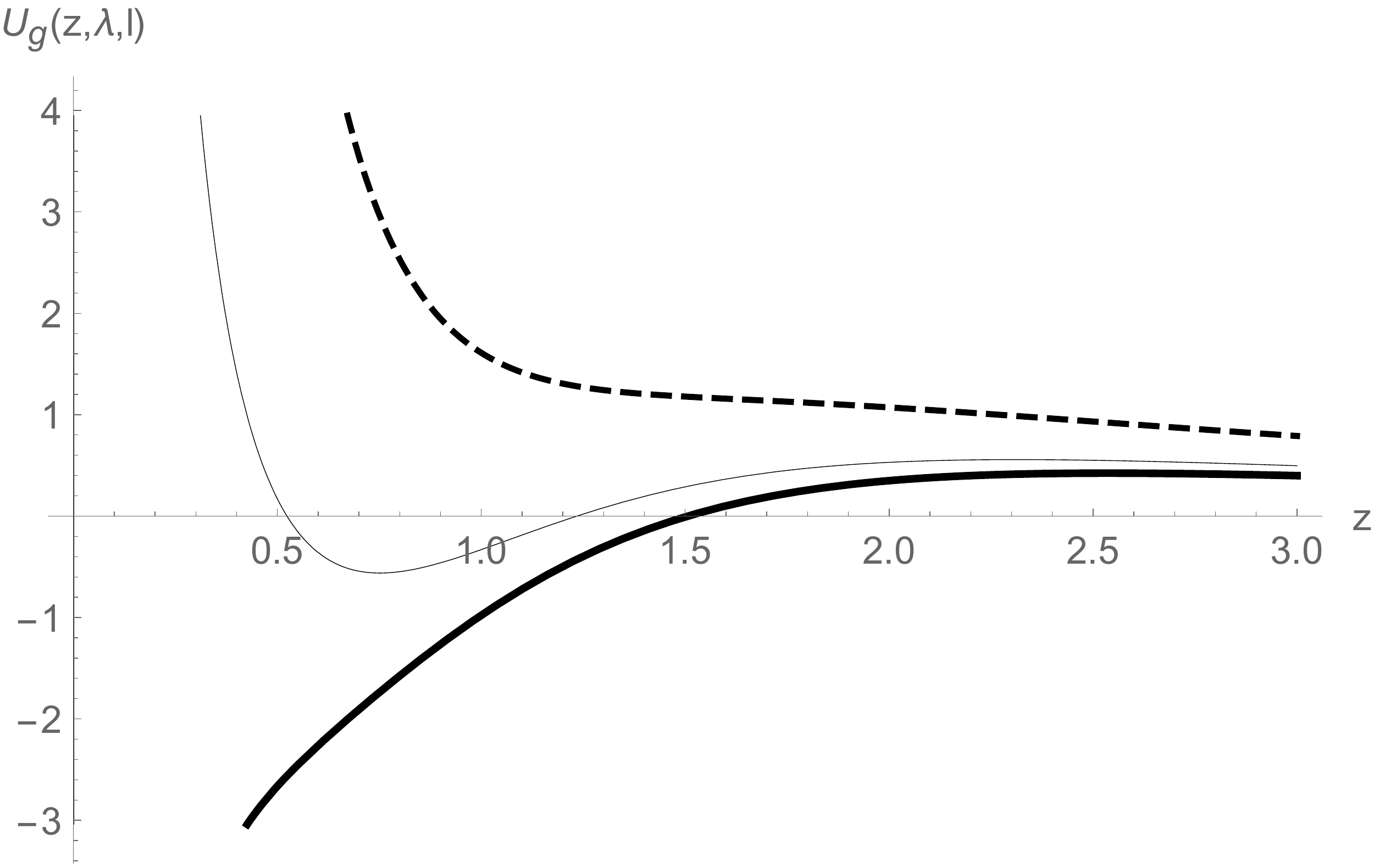}\\
           \caption{Gravitational (and scalar field) Schr\"{o}dinger potential in the warped disk model for $\lambda=0.3$ and $c=0.6$. Thick line $l=0$, thin line $l=1$ and dashed line $l=2$.}
          \label{potentialg1}
       \end{minipage}\hfill
       \begin{minipage}[b]{0.48 \linewidth}
           \includegraphics[width=\linewidth]{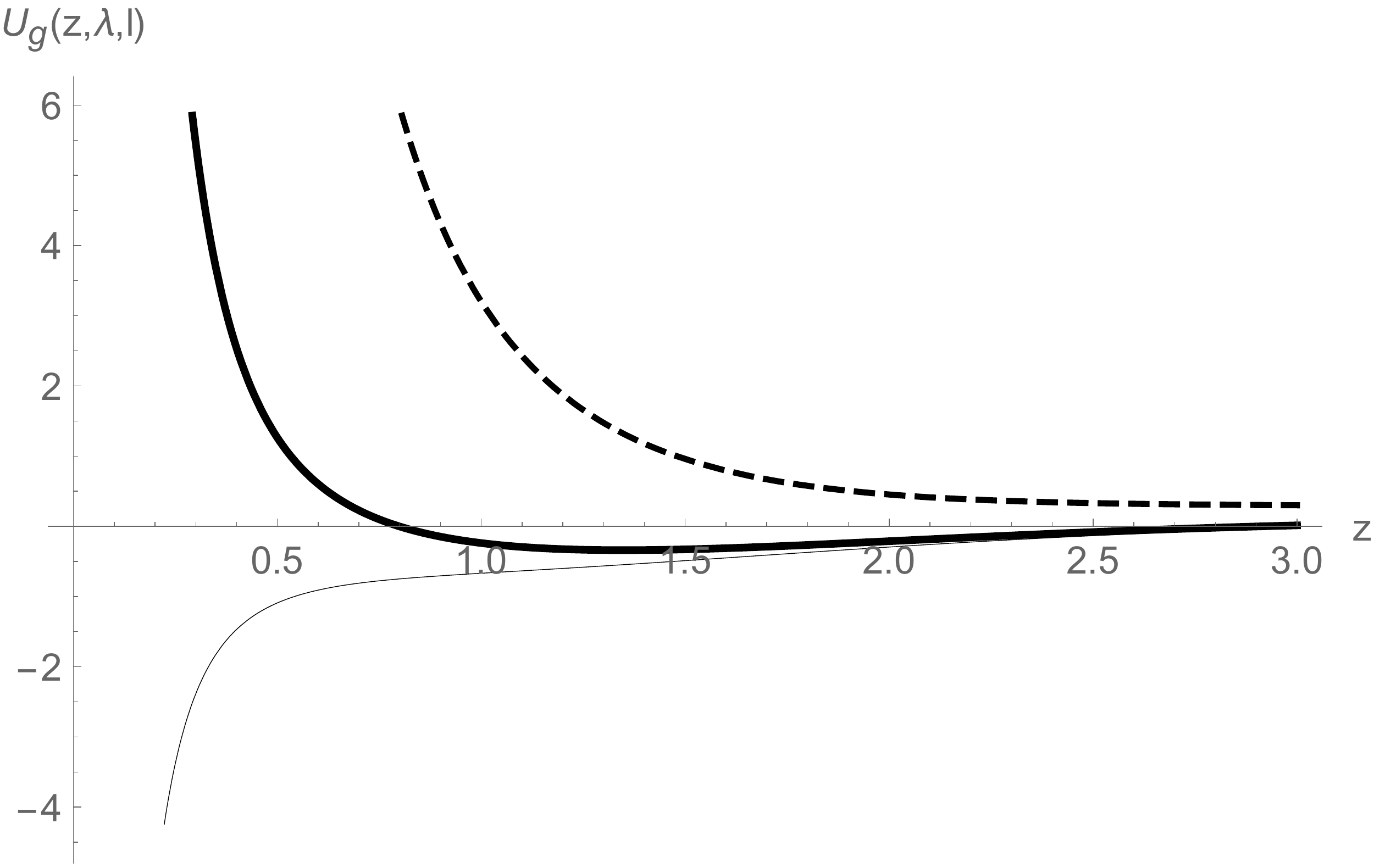}\\
           \caption{Gravitational (and scalar field) Schr\"{o}dinger potential in the exotic string model for $\lambda=0.1)$ and $c=0.6$. Thin line $(l=0)$, thick line $(l=1)$ and dashed line $l=2$.}
           \label{potencialg2}
       \end{minipage}
   \end{figure}

We plotted the graviton Schr\"{o}dinger potential for the warped disk in Fig. \ref{potentialg1} and for the exotic string in Fig. \ref{potencialg2}. Since asymptotically both potentials vanish, the massive modes are gappless free states. The scalar and vector gauge potentials have similar behavior.

The potential exhibits an infinite well at the origin for $l=0$ and an infinite barrier for $l\neq 0$. Then, only the $l=0$ states are allowed at the brane, as previously discussed. For $l=1$, it turns out that the potential has a finite well displayed from the origin, where resonant massive KK gravitons can be found.

\section{Conclusions and perspectives}
\label{cp}

We proposed a new class of smooth thick string-like model with an $AdS_6$ asymptotic regime. A localized and bell-shaped source satisfying the dominant energy condition was found, where the properties of the source and the geometry are dependent on the ratio between the cosmological constant and the brane width. A richer internal brane structure can be introduced by means of
a varying brane-tension, which modifies how the curvature and the energy density vary inside the brane core.
Amidst the thick string-like branes found, a bell-shaped source satisfying the dominant energy condition shares a
great resemblance with the numerical Abelian vortex brane \citep{Giovannini}. 

The scalar, gravitational and vector gauge sectors were also analysed. They showed similar features, as normalizable massless Kaluza-Klein modes and an attractive potential for the massive KK tower at the origin for the $l=0$ states. For $l\neq 0$, the infinite barrier at the origin avoids the detection of these modes at the brane. Nevertheless, for $l=1$, we found a potential well besides the origin, where massive resonant states could be detected.

As perspectives, we point out the use of numerical analysis to deduce these geometrical solutions from a Lagrangian model, such as a deformed Abelian vortex \cite{Giovannini}. For the KK spectrum and its phenomenological consequences, as the correction to the Newtonian and Coulomb potentials, numerical methods should also be carried out. We expect a strong influence of the brane width parameter and the bulk cosmological constant on the KK spectrum. The behavior of the massless mode and divergence the Schr\"{o}dinger potential near the brane suggests the inclusion of an interaction term between the fields and the brane, as performed in the DGP models \cite{Dvali}.

\section*{Acknowledgments}
\hspace{0.5cm}The authors would like to thank the Funda\c{c}\~{a}o Cearense de apoio ao Desenvolvimento Cient\'{\i}fico e Tecnol\'{o}gico (FUNCAP), the Coordena\c{c}\~ao de Aperfei\c{c}oamento de Pessoal de N\'ivel Superior (CAPES), and the Conselho Nacional de Desenvolvimento Cient\'{\i}fico e Tecnol\'{o}gico (CNPq) for financial support. R. V. Maluf and C. A. S. Almeida thank CNPq grant n$\textsuperscript{\underline{\scriptsize o}}$ 305678/2015-9 and 308638/2015-8 for supporting this project.


\begin{thebibliography}{99}

\bibitem{rs1}
  L.~Randall and R.~Sundrum,
  Phys.\ Rev.\ Lett.\  {\bf 83}, 3370 (1999).  

\bibitem{rs2}  
L.~Randall and R.~Sundrum,
Phys.\ Rev.\ Lett.\  {\bf 83}, 4690 (1999).

\bibitem{cosmology}
  C.~Csaki, M.~Graesser, L.~Randall and J.~Terning,
  Phys.\ Rev.\ D {\bf 62}, 045015 (2000).

\bibitem{darkmatter}
  T.~Gherghetta and B.~von Harling,
  JHEP {\bf 1004}, 039 (2010).


\bibitem{kehagias}
  A.~Kehagias and K.~Tamvakis,
Phys.\ Lett.\ B {\bf 504}, 38 (2001).

\bibitem{Bazeia}
  D.~Bazeia and A.~R.~Gomes,
  JHEP {\bf 0405}, 012 (2004).

\bibitem{dewolfe}
O.~DeWolfe, D.~Z.~Freedman, S.~S.~Gubser and A.~Karch,
Phys.\ Rev.\ D {\bf 62}, 046008 (2000).


\bibitem{gremm}
Martin Gremm, Phys. Lett. B \textbf{478}, 434-438 (2000).

\bibitem{resonance}
C. Csaki, J. Erlich, and T. J. Hollowood, 
  Phys. Rev. Lett. \textbf{84}, 5932 (2000).


\bibitem{Olasagasti}
  I.~Olasagasti and A.~Vilenkin,
  Phys.\ Rev.\ D {\bf 62}, 044014 (2000).

\bibitem{Gregory1}
  R.~Gregory,
  Phys.\ Rev.\ Lett.\  {\bf 84}, 2564 (2000).


\bibitem{Gherghetta}
  T.~Gherghetta and M.~E.~Shaposhnikov,
  Phys.\ Rev.\ Lett.\  {\bf 85}, 240 (2000).


\bibitem{Kehagias:2004fb}
 A.~Kehagias,
 Phys.\ Lett.\ B {\bf 600}, 133 (2004).



\bibitem{Cohen}
  A.~G.~Cohen and D.~B.~Kaplan,
  Phys.\ Lett.\  B {\bf 470}, 52 (1999).

\bibitem{Giovannini}
  M.~Giovannini, H.~Meyer and M.~E.~Shaposhnikov,
  Nucl.\ Phys.\ B {\bf 619}, 615 (2001).

\bibitem{Brihaye:2010nf} 
  Y.~Brihaye, T.~Delsate, N.~Sawado and Y.~Kodama,
  Phys.\ Rev.\ D {\bf 82}, 106002 (2010).

\bibitem{Oda1} I. Oda, Phys. Rev. D {\bf 62} 126009 (2000).

\bibitem{Oda2}
I. Oda,
 Phys.\ Lett.\  B {\bf 496}, 113 (2000).

\bibitem{Liufermions}
  Y.~-X.~Liu, L.~Zhao and Y.~-S.~Duan,
  JHEP {\bf 0704}, 097 (2007).




\bibitem{Tinyakov}
  P.~Tinyakov and K.~Zuleta,
  Phys.\ Rev.\ D {\bf 64}, 025022 (2001).


\bibitem{cigaruniverse}
  B.~de Carlos and J.~M.~Moreno,
  JHEP {\bf 0311}, 040 (2003).

\bibitem{stringcigar}
  J.~E.~G.~Silva, V.~Santos, C.~A.~S.~Almeida,
  Class.\ Quant.\ Grav.\  {\bf 30}, 025005 (2013).

\bibitem{resolvedconifold}
  J.~E.~G.~Silva and C.~A.~S.~Almeida,
  Phys.\ Rev.\ D {\bf 84}, 085027 (2011).

\bibitem{Torrealba}
  R.~S.~Torrealba,
  Phys.\ Rev.\ D {\bf 82}, 024034 (2010).


\bibitem{appleshapped}
  M. Gogberashvili, P. Midodashvili and D. Singleton, JHEP {\bf 0708}, 033 (2007).

\bibitem{T2}  
  Y.~-S.~Duan, Y.~-X.~Liu and Y.~-Q.~Wang,
  Mod.\ Phys.\ Lett.\ A {\bf 21}, 2019 (2006).




\bibitem{gaugecigar}
  F.~W.~V.~Costa, J.~E.~G.~Silva and C.~A.~S.~Almeida,
  Phys.\ Rev.\ D {\bf 87}, 125010 (2013).

\bibitem{Dantas}
  D.~M.~Dantas, D.~F.~S.~Veras, J.~E.~G.~Silva and C.~A.~S.~Almeida,
  Phys.\ Rev.\ D {\bf 92}, 10, 104007 (2015).





\bibitem{Dvali}
  G.~R.~Dvali, G.~Gabadadze and M.~Porrati,
  Phys.\ Lett.\ B {\bf 485}, 208 (2000).

\end{thebibliography}
\end{document}